\documentclass[a4paper,fleqn,usenatbib]{mnras}

\usepackage[T1]{fontenc}
\usepackage{ae,aecompl}

\usepackage{graphicx}	
\usepackage{amsmath}	
\usepackage{amssymb}	
\usepackage{bm}
\usepackage{color}
\usepackage{ulem}
\usepackage{cancel}
\newcommand{\bq}{\begin{equation}}
\newcommand{\eq}{\end{equation}}
\newcommand{\bqa}{\begin{eqnarray}}
\newcommand{\eqa}{\end{eqnarray}}
\newcommand{\DDt}{\frac{D}{Dt}}

\title[Convection and pulsation stability of stars]{Turbulent convection and pulsation stability of stars --
I. Basic equations for calculations of stellar structure and oscillations}

\author[D. R. Xiong et al.]{
D. R. Xiong,$^{1}$
L. Deng,$^{2}$
and C. Zhang$^{2}$
\\
$^{1}$Purple Mountain Observatory, Chinese Academy of Sciences, Nanjing 210008, China\\
$^{2}$Key Laboratory of Optical Astronomy, National Astronomical Observatories, Chinese Academy of Sciences, Beijing, 100101, China
}

\date{Accepted 2015 May 18. Received 2015 May 18; in original form 2014 December 23}

\pubyear{2015}

\begin{document}
\label{firstpage}
\pagerange{\pageref{firstpage}--\pageref{lastpage}}
\maketitle

\begin{abstract}
Starting from hydrodynamic equations, we have established a set of hydrodynamic equations for average flow and a set of dynamic equations of auto- and cross-correlations of turbulent velocity and temperature fluctuations, following the classic Reynold's treatment of turbulence. The combination of the two sets of equations leads to a complete and self-consistent mathematical expressions ready for the calculations of stellar structure and oscillations. In this paper, non-locality and anisotropy of turbulent convection are concisely presented, together with defining and calibrating of the three convection parameters ($c_1$, $c_2$ and $c_3$) included in the algorithm.  With the non-local  theory of convection, the structure of the convective envelope and the major characteristics of non-adiabatic linear oscillations are demonstrated by numerical solutions. Great effort has been exercised to the choice of convection parameters and pulsation instabilities of the models, the results of which show that within large ranges of all three parameters ($c_1$, $c_2$ and $c_3$) the main properties of pulsation stability keep unchanged.
\end{abstract}




\section{Introduction}\label{sec1}
Great advances in understanding the pulsation of variable stars have been achieved thanks to continuous work by generations of researchers for over 5 decades. Among others, we would like to name three classical books by Ledoux \& Walraven (1958), Cox (1980) and Unno et al. (1989) which reviewed and concluded exclusively the major research work done in that period of time. The theory of stellar pulsations reached a state of art stage that makes the field probably one of the bests in astrophysics. Nevertheless, some of the fundamental questions remain. Convection inside the pulsators is one of the most difficult and long-standing problems. Revolutionary observations from space by CoRoT and Kepler spaces missions, in terms of both accuracy (down to micro-magnitude) and uninterrupted time baseline (up to months), mark a new era in stellar variability studies. Thousands of small amplitude variables of many different types are discovered, which would be otherwise undetectable with ground based instruments. Such great discovery capability greatly enhanced observations of stellar variability from $\delta$ Scuti, $\gamma$ Doradus on the main sequence to pulsating red giants. Such observations provide not only opportunities but also challenges in this field, because convection replaces radiation becoming the major energy transport mechanism in stars with very extended convective envelope. As a result, dealing with the coupling between convection and oscillations is a key factor for understanding variables with low surface temperatures. However, the treatment of stellar convection, so called mixing-length theory (MLT), developed by B\"ohm-Vitense (1958) in the middle of last century is still the most popularly used in the calculations of stellar structure, evolution and oscillations. Modified versions such as time-dependent MLT (Unno 1967; Gough 1977; Stellingwerf 1982; Grigahc\`ene et al. 2005), and non-local MLT (Spiegel 1963; Ulrich 1970) had then been made in order to account for oscillations of variable stars. Needless to say that we are still far from fully understanding the nature and principal properties of stellar turbulence, therefore a robust theory of stellar convection is still missing in the community. This led to the fact that the results cannot converge  for the same observational problems when different treatments of convection were used. The theory of convection is still the most uncertain factor that prevents us from clear understandings of oscillations in low-temperature stars. In this series of papers, we are going to apply the non-local and time-dependent theory of convection we developed to the calculations of stellar oscillations. In order to cope with the great advances in observations of small amplitude red variable stars resulted from recent missions, a systematic theoretical calculations of $\delta$ Scuti, $\gamma$ Doradus and pulsating red giants are carried out, in which convection is treated by our non-local and time-dependent theory of turbulent convection. The theory was started more than 30 years ago, and has been in continuous improvements since then. Our study shows that anisotropy of turbulent convection has important effects on stellar oscillations. In the case of non-radial oscillations, neglecting anisotropy means turbulent viscosity is disregarded at the same time. Therefore the red edge of the instability strip of $\delta$ Scuti and $\gamma$ Doradus stars cannot be determined correctly. In Section~\ref{sec2}, we give a new complete version of non-local and anisotropic time-dependent convection theory and a new uniform expression for turbulent dissipation and diffusion. In Section~\ref{sec3}, we will show the general properties of the structure of stellar non-local convective envelopes. The results of theoretical calculations of non-adiabatic linear oscillations of stars are presented in Section~\ref{sec4} along with the dependance of pulsation stability on the selection of convection parameters. Conclusions and discussions are given in the last section of this paper.
\section{The basic equations}\label{sec2}
The equations used in stellar structure and oscillations can be divided into two parts: the average hydrodynamic equations, and the dynamic equations of correlations of turbulent velocity and temperature. Both of them come from the literal hydrodynamic equations, as in the following,

\bq
\frac{\partial \left(\rho u^i\right)}{\partial t}+\nabla_k\left(\bar\rho u^i u^k+g^{ik}P\right)+\bar\rho g^{ik}\nabla_k\phi=\nabla_k\sigma^{ik}\left( u\right),\label{eq1}
\eq

\bq
\frac{\partial\rho}{\partial t}+\nabla_k\left(\rho u^k\right)=0,\label{eq2}
\eq

\bq
\rho T\left(\frac{\partial S}{\partial t}+u^k\nabla_kS\right)=\rho\epsilon_N+\sigma^{ik}\left( u\right)\nabla_k u_i-\nabla_kF^k_\mathrm{r},\label{eq3}
\eq

 \bq
 F^k_\mathrm{r}=-\frac{4acT^3}{3\rho\kappa}g^{ik}\nabla_kT,\label{eq4}
 \eq

 \bq
 g^{\alpha\beta}\nabla_\alpha\nabla_\beta\phi=4\pi G\rho,\label{eq5}
 \eq

 \noindent where $\rho$, $T$ and $P$ are the density, temperature and pressure of gas,  $S$, $\epsilon_N$ and $\kappa$ are the entropy of gas, nuclear energy general rate and radiative opacity, respectively. $u^i$ is the $i$th component of the velocity of fluid in the curved framework of reference, $\sigma^{ik}$ is the viscous stress tensor, $F^k_\mathrm{r}$ is the the radiation flux vector, and $\phi$ is the gravitational potential. Equations (\ref{eq1})--(\ref{eq3}) are the equations for the conservations of momentum, mass and energy, while equation (\ref{eq4}) is that of radiative transfer, equation (\ref{eq5}) is the Poisson's equation for self-gravitating system. The equations are all expressed in tensor format, in which the implicit summation rule of tensor is used, i.e. a summation should be performed for a index running from 1 to 3, $i$ for instance, if a pair of sub- and super-script of that index appear in a term.  $g^{\alpha\beta}$ is the metric tensor of the curved framework of reference.

\subsection{The average hydrodynamic equations}\label{sec21}
In our formalism of convection theory, the classical Reynold's algorithm is adopted. Due to the vast physical dimension of fluid in stars, turbulent convection is bound to happen once convective motion starts. In Reynold's scheme, each of the physical quantities $X$ (gas density, temperature, pressure, entropy etc.) is expressed as its averaged value $\bar X$ plus the corresponding fluctuation $X'$ (apply in all of our earlier work, Xiong 1978, 1989; Xiong et al. 1997),

\bq
X=\bar{X}+X',
\label{eq6}
\eq

Applying the form of equation (\ref{eq6}) to equations (\ref{eq1})--(\ref{eq5}), making Taylor expansion of $X'$ and retaining only its 1st-order terms, then averaging each of the equations, we have the following average hydrodynamic equations (Xiong 1978, 1989; Xiong et al. 1997),

\bq
\frac{D\overline{u^i}}{Dt}+{1\over{\bar\rho}}\nabla_k\left(g^{ik}\bar P+\overline{\rho u'^iu'^k}\right)+g^{ik}\nabla_k\bar{\phi}={1\over{\bar\rho}}\nabla_k\sigma^{ik}\left(\bar u\right),\label{eq7}
\eq

\bq
\frac{D\bar\rho}{Dt}+\bar\rho\nabla_k\overline{u^k},\label{eq8}
\eq

\bq
\bar\rho{\bar C}_P\frac{D\bar T}{Dt}-\alpha\frac{D\bar P}{Dt}+\nabla_k\left(\bar\rho{\bar C}_P\bar T\overline{w'^kT'/\bar T}\right)+\nabla_k\overline{F^k_\mathrm{r}}=\bar\rho\bar{\epsilon}_N+\overline{\sigma^{ik}\left( u\right)\nabla_ku_i},\label{eq9}
\eq

\bq
\overline{F^k_\mathrm{r}}= -\frac{4ac\bar T^3}{3\bar\rho\bar\kappa}g^{ik}\nabla_k\bar T,\label{eq10}
\eq

\bq
g^{\alpha\beta}\nabla_\alpha\nabla_\beta\bar\phi=4\pi G\bar\rho,\label{eq11}
\eq

\noindent where,

\bq
\frac{D}{Dt}=\frac{\partial}{\partial t}+u^k\nabla_k, \label{eq12}
\eq

\noindent is the co-moving differential, $C_P$ is specific heat at constant pressure, $\alpha=-\left(\partial\ln\rho/\partial\ln T\right)_P$ is the expansion coefficient of gas.

The average form of hydrodynamic equations (equations (\ref{eq7})--(\ref{eq11})) is similar as their original form (equations (\ref{eq1})--(\ref{eq5})). When convection sets in, however, an extra term, $\overline{\rho u'^iu'^k}$ emerges in the average equation of momentum conservation (equation (\ref{eq7})), which is nothing but the well known Reynold's stress. And another extra term, $\bar\rho{\bar C}_P\bar T\overline{w'^kT'/\bar T}$  appears in the equation of energy conservation (equation (\ref{eq9})), which is convective enthalpy flux. These two terms represent respectively momentum and thermal energy transport caused by convection in the fluid. The next step in our theory is to formulate the dynamic equations for these terms. Once the dynamic equations of the second-order correlation of turbulent velocity and temperature are established, the whole set of dynamic equations of turbulent convection becomes solvable by combination of resulted equations with those of averaged dynamic equations (equations (\ref{eq7})--(\ref{eq11})).

\subsection{Dynamic equations of turbulent velocity and temperature}\label{sec22}

Subtraction of the average equations (\ref{eq7})--(\ref{eq9}) from each other's corresponding original ones equations (\ref{eq1})--(\ref{eq3}), it is trivial to have the following dynamic equations of turbulent velocity and temperature,

\bqa
\frac{Dw'^i}{Dt}+w'^k\nabla_k\overline{u^i}+{1\over\bar\rho}\nabla_k\left(g^{ik}P'+\rho u'^iu'^k-\overline{\rho u'^iu'^k}\right)\nonumber\\
-\alpha\frac{T'}{\bar T}\left(g^{ik}\nabla_k\bar\phi+\frac{D\overline{u^i}}{Dt}\right)={1\over\bar\rho}\nabla_k\sigma^{ik}\left(u'\right),
\label{eq13}
\eqa

\bqa
\lefteqn{\DDt\frac{T'}{\bar T}+\frac{T'}{\bar T}\left\{\left[1-\bar\alpha+{\bar C}_{P,T}\right]\frac{D\ln\bar T}{Dt}\right.}\nonumber\\
& & \mbox{}+\left. \left[{\bar C}_{P,P}+\left(1-\bar\alpha\right){\bar\nabla}_\mathrm{ad}\right]\frac{D\ln\bar P}{Dt}\right\}\nonumber\\
& & \mbox{}+w'^k\left(\nabla_k\ln\bar T-{\bar\nabla}_\mathrm{ad}\nabla_k\ln\bar P\right)\nonumber\\
& & \mbox{}+\frac{1}{\bar\rho{\bar C}_P\bar T}\nabla_k\left[\bar\rho\bar{C}_P\bar T\left(w'^k\frac{T'}{\bar T}-\overline{w'^k\frac{T'}{\bar T}}\right)\right]\nonumber\\
& & \mbox{} =\frac{1}{\bar\rho{\bar C}_P\bar T}\left[\left(\rho\epsilon_N\right)'+\sigma^{ik}\left( u\right)\nabla_ku_i-\overline{\sigma^{ik}\left( u\right)\nabla_ku_i}-\nabla_kF'^k_\mathrm{r}\right],\nonumber\\
\label{eq14}
\eqa

\noindent where $w'^i=\rho u'^i/\bar\rho$ is the density-weighted turbulent velocity, $\nabla_\mathrm{ad}=\alpha P/\rho C_PT=(\Gamma_2-1)/\Gamma_2$ is adiabatic temperature gradient, ${\bar C}_{P,T}$ and ${\bar C}_{P,P}$ are respectively the partial derivatives of ${\bar C}_P$ with respect to $T$ and $P$.

\subsection{Dynamic equations of correlations of turbulent velocity and temperature}\label{sec23}

Starting from equations (\ref{eq13}) and (\ref{eq14}), it is rather straightforward to make the dynamic equations of the auto- and cross-correlations of turbulent velocity and temperature,

\bqa
\lefteqn{\DDt{\overline{w'^iw'^j}}+\overline{w'^iw'^k}\nabla_k\overline{u^j}}\nonumber\\
& & \mbox{}+\overline{w'^jw'^k}\nabla_k\overline{u^i}+\frac{1}{\bar\rho}\nabla_k\left(\bar\rho\overline{u'^kw'^iw'^j}\right)\nonumber\\
& & \mbox{}-\overline{w'^i\frac{T'}{\bar T}}\left(\frac{D\overline{u^j}}{Dt}+g^{jk}\nabla_k\bar\phi\right)-\overline{w'^j\frac{T'}{\bar T}}\left(\frac{D\overline{u^i}}{Dt}+g^{ik}\nabla_k\bar\phi\right)\nonumber\\
& &  \mbox{}+{1\over\bar\rho}\nabla_k\left[g^{ik}\overline{w'^jp'}+g^{jk}\overline{w'^ip'}-\overline{w'^i\sigma^{jk}\left(u'\right)}+\overline{w'^j\sigma^{ik}\left(u'\right)}\right]\nonumber\\
& & \mbox{}-{1\over\bar\rho}\left(g^{ik}\overline{P'\nabla_kw'^j}-g^{jk}\overline{P'\nabla_kw'^i}\right)\nonumber\\
& & \mbox{}= -{1\over\bar\rho}\left[\overline{\sigma^{ik}\left(u'\right)\nabla_kw'^j}+\overline{\sigma^{jk}\left(u'\right)\nabla_ku'^i}\right],
\label{eq15}
\eqa

\bqa
\lefteqn{\DDt{\overline{\left(\frac{T'}{\bar T}\right)^2}} +2\overline{\left(\frac{T'}{\bar T}\right)^2}}\nonumber\\
& & \mbox{} \times\left\{\left[1-\bar\alpha+{\bar C}_{P,T}\right]\frac{D\ln\bar T}{Dt}+
\left[{\bar C}_{P,P}+\left(1-\bar\alpha\right)\bar\nabla_\mathrm{ad}\right]\frac{D\ln\bar P}{Dt}\right\}\nonumber\\
& & \mbox{} +2\overline{w'^k\frac{T'}{\bar T}}\left(\nabla_k\ln\bar T-\bar\nabla_\mathrm{ad}\nabla_k\ln\bar P\right)\nonumber\\
& & \mbox{} +\frac{1}{\bar\rho^2\bar{C}_P^2}\nabla_k\left[\bar\rho^2\bar{C}_P^2\overline{w'^k\left(\frac{T'}{\bar T}\right)^2}\right]\nonumber\\
& & \mbox{} -\overline{\frac{2}{\bar\rho\bar{C}_P\bar T}\frac{T'}{\bar T}\left(\rho\epsilon_N\right)'}=-\frac{2}{\bar\rho\bar{C}_P\bar T}\overline{\frac{T'}{\bar T}\nabla_kF'^k_\mathrm{r}},
\label{eq16}
\eqa

\bqa
\lefteqn{\DDt\overline{w'^i\frac{T'}{\bar T}}+\overline{w'^k\frac{T'}{\bar T}}\nabla_k\overline{u'^i}+\overline{w'^i\frac{T'}{\bar T}}}\nonumber\\
& &\mbox{}\times\left\{\left[1-\bar\alpha+\bar{C}_{P,T}\right]\frac{D\ln\bar T}{Dt}+\left[{\bar C}_{P,P}+\left(1-\bar\alpha\right)\bar\nabla_\mathrm{ad}\right]\frac{D\ln\bar P}{Dt}\right\}\nonumber\\
& & \mbox{} -\bar\alpha\overline{\left(\frac{T'}{\bar T}\right)^2}\left(g^{ik}\nabla_k\bar\phi+\frac{D\overline{u^i}}{Dt}\right)\nonumber\\
& & \mbox{} +\overline{w'^iw'^k}\left(\nabla_k\ln\bar T-\bar\nabla_\mathrm{ad}\nabla_k\ln\bar P\right)\nonumber\\
& & \mbox{} +\frac{1}{\bar\rho\bar{C}_P}\nabla_k\left[\bar\rho\bar{C}_P\overline{u'^kw'^i\frac{T'}{\bar T}}\right]\nonumber\\
& & \mbox{} =\frac{1}{\bar\rho}\overline{\frac{T'}{\bar T}\nabla_k\sigma^{ik}\left(u'\right)}-\frac{1}{\bar\rho\bar{C}_P\bar T}\overline{w'^i\nabla_kF'^k_\mathrm{r}},
\label{eq17}
\eqa

Convection is an intrinsic instability induced by thermal instability in fluid. Equations (\ref{eq15})--(\ref{eq17}) are the dynamic equations describing the auto- and cross-correlations of turbulent velocity and temperature in stars. Taking equation (\ref{eq15}) as an example, it can be regarded as the conservation of turbulent kinetic energy, the first term in which represents the variation rate of turbulent kinetic energy, therefore is the sum of all the other terms with a negative sign, while the second and the third terms are the exchange rates between the average kinetic energy and turbulent energy due to the shear force of average motion of the fluid; the 5th and 6th terms are the variation rates of turbulent kinetic energy caused by buoyancy force; the 4th term is the pure gain of turbulent kinetic energy due to momentum exchange of non-local turbulent kinetic energy flux (defined as $\overline{\rho u'^kw'^iw'^j}$); the brackets in the 7th term are the pressure tensor flux and viscous tensor flux; the last two terms in the left-hand side and the terms in the right-hand side in equation (\ref{eq15}) will be discussed in detail later.

\subsection{Turbulent dissipation}\label{sec24}

The right-hand side terms in equations (\ref{eq15})--(\ref{eq17}) represent attenuations caused by molecular viscosity and radiation conductivity. Following the theory of turbulence (Hinze 1975), these terms can be expressed as (see Xiong 1978),

\bq
{1\over\bar\rho}\left[\overline{\sigma^{ik}\left(u'\right)\nabla_ku'^j}+\overline{\sigma^{jk}\left(u'\right)\nabla_ku'^i}\right]=-2\sqrt{3}\eta_ex\overline{w'^iw'^j}/l_e,
\label{eq18}
\eq

\bq
-\frac{1}{\bar\rho\bar{C}_P\bar T}\overline{\frac{T'}{\bar T}\nabla_kF'^k_\mathrm{r}}=-\sqrt{3}\eta_ex\left(1+\frac{x_c}{x}\right)\overline{\left(\frac{T'}{\bar T}\right)^2}/l_e,
\label{eq19}
\eq

\bq
{1\over\bar\rho}\overline{\frac{T'}{\bar T}\nabla_k\sigma^{ik}\left(u'\right)}-\frac{1}{\bar\rho\bar{C}_P\bar T}\overline{w'^i\nabla_kF'^k_\mathrm{r}}=-\sqrt{3}\eta_ex\left(3+\frac{x_c}{x}\right)\overline{w'^i\frac{T'}{\bar T}}/l_e
\label{eq20}
\eq

\noindent where $\eta_e=0.45$ is the Heisenberg eddy coupling constant (Hinze 1975), $x$ and $x_c/x$ are respectively the rms turbulent velocity and the effective Peclet number, $l_e$ is the length of energy-containing eddies of turbulence, defined as,

\bq
x=\sqrt{g_{ij}\overline{w'^iw'^j}}/3,\label{eq21}
\eq

\bq
x_c=3acGM_r\bar T^3/c_1\bar\rho\bar\kappa\bar{C}_Pr^2\bar P,
\label{eq22}
\eq
where $c_1$ is a convective parameter (see equation (\ref{eq31})).

Non-locality and anisotropy of turbulence are the most fundamental and important properties of stellar convection, both of which play key roles in stellar evolution and pulsation instability and will be discussed in sections~\ref{sec25} and \ref{sec26}.

\subsection{Non-locality of turbulent convection}\label{sec25}

The original mixing-length theory (B\"ohm-Vitense 1958) is a phenomenological and local expression of stellar convection. As pointed out by Speigel (1963), the local mixing-length theory would make sense only when convective motion were homogeneous in an infinitely large field, or when the so called mixing-length was far smaller than the characteristic dimension of fluid media. Unfortunately, these two length scales are usually comparable in stellar interiors,  and the temperature gradient generally changes considerably within a mixing-length in stellar fluid media. To overcome such a shortcoming, one can replace the local temperature gradient at a given point in the fluid by the averaged value over a mixing-length (Speigel 1963). In such a way, the general treatment of convection became a non-local mixing-length theory.

Our theory of stellar convection described above is originated from hydrodynamic equations and turbulent theory, therefore it naturally covers non-locality of convective motion in stars. In the equations of correlations,  equations (\ref{eq15})--(\ref{eq17}), the third-order terms represent non-locality of turbulent convection. In fact, the term $\overline{\rho u'^kw'^iw'^j}$ is turbulent kinetic energy flux which describes non-local transportation of turbulent energy from a point in the fluid to another. In the second-order dynamic equations for the correlation of  turbulent velocity and temperature fluctuations, the third-order terms are bound to emerge, while in the third-order equations the fourth-order terms must turn up, and the loop keeps going for any order of correlations. This is the well-known closure problem in the study of turbulence. A possible approach to such problem we provided is a gradient-type diffusive approximation, i.e. assuming,

\bq
\overline{\rho u'^kw'^iw'^j}=-\bar\rho\overline{w'^kw'^\alpha}\tau_c\nabla_\alpha\overline{w'^iw'^j},
\label{eq23}
\eq
\bq
\overline{u'^k\left(\frac{T'}{\bar T}\right)^2}=-\overline{w'^kw'^\alpha}\tau_c\nabla_\alpha\overline{\left(\frac{T'}{\bar T}\right)^2},
\label{eq24}
\eq
\bq
\overline{u'^kw'^i\frac{T'}{\bar T}}=-\overline{w'^kw'^\alpha}\tau_c\nabla_\alpha\overline{w'^i\frac{T'}{\bar T}},
\label{eq25}
\eq

\noindent where,

\bq \tau_c=\Lambda/x,\label{eq26}\eq

\noindent is the lifetime of a turbulent eddy, $\Lambda$ is the corresponding diffusion length scale.

Another way of closure of the correlation equations is to formulate the dynamic equations of the third-order terms, and to deal with the fourth-order correlations that emerge in the process. Following the theory of stochastic processes, a fourth-order correlation can be represented by the product of two second-order correlations, given the stochastic quantities obey Gaussian distribution. Grossman (1996) made a very thorough comparison between the two ways of closure, and he concluded that for the second-order correlations, the former algorithm (equations (\ref{eq23})--(\ref{eq25})) is more preferable. For the calculations of stellar structure and oscillations, the second order correlations are more important. For that reason, he further pointed out that the closure we offered is also more practical.

\subsection{Anisotropy of turbulent convection}\label{sec26}

The study of turbulence has a long history and is still under development. All the existing theories are based on an isotropic assumption. For stellar convection, however, such an assumption obviously breaks down. Observations of the velocity field of solar granulation show that the convective motions are primarily radial in the unstable convective region under the photosphere, while they turn to be primarily horizontal in the convective overshooting zone in the outer layers of atmosphere. Such a definitive evidence requires a theory that accounts for anisotropy of turbulence when dealing with convective overshoot problem. Anisotropy of turbulence reduces the efficiency of overshooting mixing. Our studies demonstrated that the dynamic coupling between convection and oscillations plays a key role for the oscillation instability in low-temperature stars with extended convective envelopes, and anisotropy of turbulence is ultimately linked to turbulent viscosity. The dynamic theory of correlations of turbulent convention is based on hydrodynamics and theory of turbulence, therefore it is enabled to handle anisotropic turbulent convection in stars. We also generalized the treatment of anisotropy by Canuto (1993) for a more robust application, decomposing Renold's stress $\overline{w'^iw'^j}$ into the sum of an isotropic component $g^{ij}x^2$ and an anisotropic one $\chi^{ij}$,

\bq
\overline{w'^iw'^j}=g^{ij}x^2+\chi^{ij},\label{eq27}
\eq

\noindent where,

\bq
x^2=g_{ij}\overline{w'^iw'^j}/3.\label{eq28}
\eq

From equations (\ref{eq27})--(\ref{eq28}), it is clear that,

\bq
g_{ij}\chi^{ij}=0,\label{eq29}
\eq

Pressure fluctuation terms are also present in the dynamic equations of turbulent velocity and temperature correlations. From the theory of turbulence, we know that the correlation of pressure and velocity gradient tends to turn turbulence into isotropic (Rotta 1951, Hinze 1975), therefore we assume,

\bq
{1\over\bar\rho}\overline{P'\left(g^{ik}\nabla_kw'^j+g^{jk}\nabla_kw'^i-{2\over 3}g^{ij}\nabla_ku'^k\right)}= -c_3\frac{4\sqrt{3}\eta_e}{3}x\chi^{ij}/l_e,\label{eq30}
\eq

\noindent where $c_3$ is the parameter used to describe the anisotropy of turbulent convection. The larger $c_3$, the stronger the tendency for turbulent pressure fluctuations to restore isotropy, therefore turbulence is made more isotropic, and vise-versa. In our theory, the ratio of the squared radial component of turbulent velocity to that of horizontal motion is $\overline{w'^2_r}/\overline{w'^2_h}=\mathbf{\left(3+c_3\right)/2c_3}$ in the unstable region (Deng et al. 2006).

As shown in equations (\ref{eq18})--(\ref{eq20}) and (\ref{eq24})--(\ref{eq26}) that $l_e$ and $\Lambda$ are two characteristic lengths that are related to turbulent dissipation and diffusion. We then further assume that they are both proportional to the local pressure scale height $H_P$,

\bq
l_e=c_1H_P=c_1\frac{r^2\bar P}{GM_r\bar\rho},\label{eq31}
\eq

\bq
\Lambda=\frac{\sqrt{3}}{4}c_2H_P=c_2\frac{\sqrt{3}r^2\bar P}{4GM_r\bar\rho},\label{eq32}
\eq

Inserting equations (\ref{eq27})--(\ref{eq30}), together with equations (\ref{eq18})--(\ref{eq20}) that describe turbulent dissipation and equations (\ref{eq23})--(\ref{eq26}) that describe turbulent diffusion into equations (\ref{eq7})--(\ref{eq11}) and equations (\ref{eq15})--(\ref{eq17}), also applying above assumptions equations (\ref{eq31})--(\ref{eq32}), the dynamic equations for the calculation of stellar structure and oscillations in non-local and anisotropic framework can then be re-written as the following,

\bq
\frac{D\overline{u^i}}{Dt}+{1\over \bar\rho}\nabla_k\left[g^{ik}\left(\bar P+\bar\rho x^2\right)+\bar\rho\chi^{ik}\right]+g^{ik}\nabla_k\bar\phi=0,
\label{eq33}
\eq

\bq
\frac{D\bar\rho}{Dt}+\bar\rho\nabla_k\overline{u^k}=0,
\label{eq34}
\eq

\bqa
\frac{D\ln\bar T}{Dt}-\bar\nabla_\mathrm{ad}\frac{D\ln\bar P}{Dt}+\frac{1}{\bar\rho\bar C_P\bar T}\left[\bar\rho x^2\left(3\frac{D\ln x}{Dt}-\frac{D\ln\bar\rho}{Dt}\right)\right.\nonumber\\
\left.+\bar\rho\chi^{ik}\nabla_k\overline{u_i}+\nabla_k\left(\overline{F_\mathrm{r}^k}+\overline{F_\mathrm{c}^k}+\overline{F_\mathrm{t}^k}\right)\right]=0,
\label{eq35}
\eqa

\bq
\overline{F^k_\mathrm{r}}=-\frac{4ac\bar T^3}{3\bar\rho\bar\kappa}g^{ik}\nabla_i\bar T,
\label{eq36}
\eq

\bq
g^{\alpha\beta}\nabla_\alpha\nabla_\beta\bar\phi=4\pi G\bar\rho,
\label{eq37}
\eq

\bqa
{3\over 2}\frac{Dx^2}{Dt}-x^2\frac{\ln\bar\rho}{Dt}+\chi^{ik}\nabla_i\overline{u_k}-\frac{3}{2\bar\rho}\nabla_i\left(Q^{ik}\nabla_kx^2\right)\nonumber\\
-\bar\alpha V^k\left(\frac{D\overline{u_k}}{Dt}+\nabla_k\bar\phi\right)=-\frac{2\sqrt{3}\eta_eGM_r\bar\rho}{c_1r^2\bar P}x^3,
\label{eq38}
\eqa

\bqa
\lefteqn{\frac{D\chi^{ij}}{Dt}+x^2\left(g^{ik}\nabla_k\overline{u^i}+g^{jk}\nabla_k\overline{u^i}-{2\over 3}g^{ij}\nabla_k\overline{u^k}\right)}\nonumber\\
&&\mbox{} +\chi^{ik}\nabla_k\overline{u^j}+\chi^{jk}\nabla_k\overline{u^i}-{2\over 3}g^{ij}\chi^{\alpha\beta}\nabla_\alpha\overline{u_\beta}\nonumber\\
&&\mbox{}-{1\over\bar\rho}\nabla_\alpha\left(Q^{\alpha k}\nabla_k\chi^{ij}\right)\nonumber\\
&&\mbox{}-\bar\alpha\left(g^{ik}V^j+g^{jk}V^i-{2\over 3}g^{ij}V^k\right)\left(\frac{D\overline{u_k}}{Dt}+\nabla_k\bar\phi\right)\nonumber\\
&&\mbox{}=-\frac{4\sqrt{3}\eta_eGM_r\bar\rho\left(1+c_3\right)}{3c_1r^2\bar P}\chi^{ij}
\label{eq39}
\eqa

\bqa
\lefteqn{\frac{DZ}{Dt}+Z\left\{\left[1-\bar\alpha+{\bar C}_{P,T}\right]\frac{D\ln\bar T}{Dt}\right.}\nonumber\\
&&\mbox{}+\left.\left[\bar C_{P,P}+\left(1-\bar\alpha\right)\bar\nabla_\mathrm{ad}\right]\frac{D\ln\bar P}{Dt}\right\}\nonumber\\
&&\mbox{}+2V^k\left(\nabla_k\ln\bar T-\bar\nabla_\mathrm{ad}\ln\bar P\right)\nonumber\\
&&\mbox{}-\frac{1}{\bar\rho^2\bar C_P^2}\nabla_k\left(\bar\rho\bar C_P^2Q^{ik}\nabla_kZ\right)\nonumber\\
&&\mbox{}=-\frac{2\sqrt{3}\eta_eGM_r\bar\rho}{c_1r^2\bar P}\left(x+x_c\right)Z
\label{eq40}
\eqa

\bqa
\lefteqn{\frac{DV^i}{Dt}+V^k\nabla_k\overline{u^i}+V^i\left\{\left[1-\bar\alpha+\bar C_{P,T}\right]\frac{D\ln\bar T}{Dt}\right.}\nonumber\\
&&\mbox{}\left.+\left[\bar C_{P,P}+\left(1-\bar\alpha\right)\bar\nabla_\mathrm{ad}\right]\frac{D\ln\bar P}{Dt}\right\}\nonumber\\
&&\mbox{}-\alpha Z\left(\frac{D\overline{u^i}}{Dt}+g^{ik}\nabla_k\bar\phi\right)\nonumber\\
&&\mbox{}+\left(g^{ik}x^2+\chi^{ik}\right)\left(\nabla_k\ln\bar T-\nabla_k\ln\bar P\right)\nonumber\\
&&\mbox{}-\frac{1}{\bar\rho\bar C_P}\nabla_i\left(\bar C_PQ^{ik}\nabla_kV^i\right)\nonumber\\
&&\mbox{}=-\frac{\sqrt{3}\eta_eGM_r\bar\rho}{c_1r^2\bar P}\left(3x+x_c\right)V^i,
\label{eq41}
\eqa

\noindent where,

\bq
Z=\overline{\left(\frac{\bar T}{T'}\right)^2},
\label{eq42}
\eq

\bq
V^i=\overline{w'^i\frac{T'}{\bar T}},
\label{eq43}
\eq

\bq
\overline{F_\mathrm{c}^k}=\bar\rho\bar C_P\bar TV^k,
\label{eq44}
\eq

\bq
\overline{F_\mathrm{t}^k}=-Q^{ik}\nabla_kx,
\label{eq45}
\eq

\bq
Q^{ij}=\frac{\sqrt{3}\pi c_2r^2\bar Px}{GM_r}\left(g^{ij}+\chi^{ij}/x^2\right),
\label{eq46}
\eq

Now, equations (\ref{eq33})--(\ref{eq41}) form a complete set of dynamic equations for the calculation of stellar structure and oscillations.

\subsection{Calibrations of convective parameters $c_1$, $c_2$ and $c_3$}\label{sec27}

After proper treatment for turbulent dissipation, diffusion and anisotropy, the original equations (\ref{eq15})--(\ref{eq17}) have been transformed to the dynamic equations of non-local and anisotropic convection (\ref{eq38})--(\ref{eq41}). Three parameters ($c_1$, $c_2$ and $c_3$) are included during parameterization of turbulent dissipation (equations (\ref{eq18})--(\ref{eq20})), turbulent diffusion (equations (\ref{eq23})--(\ref{eq25})) and anisotropy of turbulent convection (equation (\ref{eq30})). They should be a function varying slowly with mass $M$, luminosity $L$, and effective temperature $T_\mathrm{e}$ of stars (Ludwig et al. 1999). To date, the most reliable way for the calibrations of parameters introduced in different types of convection theories is on the basis of the Sun. Limited observational constraints can hardly calibrate separately the three parameters in our theory. $c_3$ is rather an independent parameter which can, in principle, be calibrated using the variations of radial and horizontal turbulent velocity and temperature as functions of depth in solar outer layers. Unfortunately, observations capable of doing such a calibration are not yet accurate enough to ensure good calibrations of $c_3$. From stellar evolution point of view, $c_3$ is more linked to the structure of convective overshooting zones in stars.  There is convergent evidence from observations of solar granulation velocity, hydrodynamic simulations and numerical calculations using our non-local and anisotropic convection theory, that are all inclined to the same conclusion: turbulent motion is primarily radial in convectively unstable zone, while it becomes more and more dominant in horizontal motion when going into overshooting region. The major function of anisotropy of turbulence there is to reduce the depth of overshooting distance. Fortunately, the ratio of the radial and horizontal components ($u'_r$ and $u'_h$) of turbulent velocity is approximately $\overline{u'^2_r}/\overline{u'^2_h}\approx1/2$ in convective overshooting zone, which is almost independent of the choice of $c_3$. Many studies, including observations of the velocity field of solar granulations, hydrodynamic simulations (Deng et al. 2006) and calculations of lithium abundance profile in the Sun and stars (Xiong \& Deng 2009), show that $c_3\approx 3$ is a plausible value. That infers that $\overline{u'^2_r}/\overline{u'^2_h}\approx 1$ in the convectively unstable zone, and it is also in good agreement with the most unstable modes from linear oscillation stability analysis of convective modes (Unno 1961).

As $c_1$ and $c_2$ are the two convection parameters that represent the characteristic lengths $l_e$ and $\Lambda$ of turbulence (equations (\ref{eq31}) and (\ref{eq32})), therefore it is reasonable to believe that the ratio of the two parameters $c_2/c_1$ should be close to a constant of order unity. As a result, once $c_3$ and $c_2/c_1$ are fixed, $c_1$ becomes the only one adjustable parameter in the mathematical scheme of our theory.

Based on systematic analysis of all the results from observations of the velocity field of solar granulations (Keil \& Canfield 1978; Neisis \& Mattig 1989; Komm et al. 1991), assessment of the surface Lithium abundance in late-type dwarf stars, using hydrodynamic simulations, and the comparison between adiabatic helioseismic inversion and the theoretical models (Zhang et al. 2012), we have come to a set of parameters ($c_1$, $c_2/c_1$, $c_3$)=(0.64, 0.50, 3) that is robust in dealing with stellar convection.

\section{The structure of convective envelope of stars}\label{sec3}

Equations suitable for the structure of stellar envelopes in non-local convection theory can be derived by putting all terms containing velocity $\bar u$ and its time-derivatives in equations (\ref{eq33})--(\ref{eq41}). For the calculations of equilibrium stellar model, equation (\ref{eq37}) has a first integration,

\bq
\frac{\partial\phi}{\partial r}=\frac{GM_r}{r^2},
\label{eq47}
\eq

\begin{figure}
\includegraphics[width=\columnwidth]{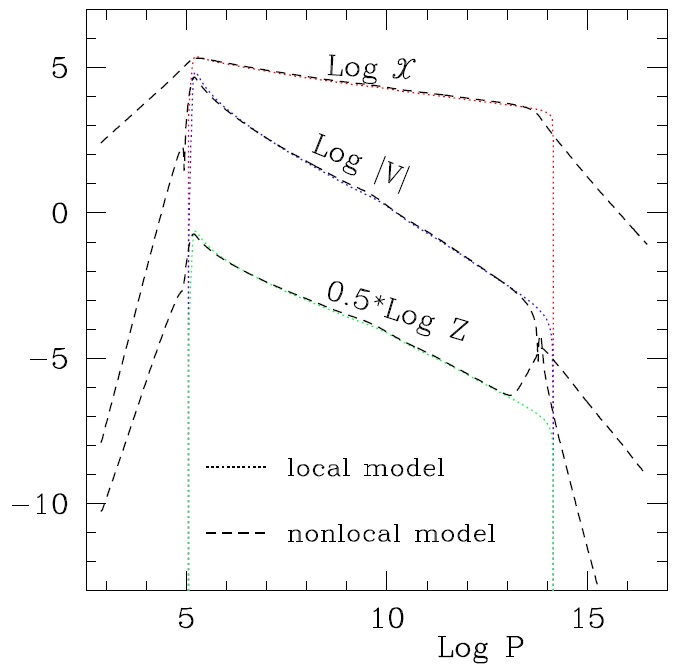}
\caption{The correlations $x$, $Z$, $V$ of turbulent velocity and temperatures versus $\log P$ for the local (dotted lines) and non-local (dashed lines) models of the Sun.}\label{fig1}
\end{figure}

Then, inserting equation (\ref{eq47}) into (\ref{eq33}), equation (\ref{eq37}) can be eliminated. $\phi$ not anymore shows up in equations (\ref{eq33})--(\ref{eq36}) and (\ref{eq38})--(\ref{eq41}). In such a way, the whole equations degrade two orders. The number of equations for equilibrium envelope structure is now 12. For the boundary conditions and numerical scheme of working equations in non-local convection, please refer to our previous work (Deng \& Xiong 2008). It is emphasized here that convective envelope structure in local condition is only an initial value problem of solving the 4th-order differential equations, which can be integrated from stellar surface down to the bottom of convective zone.  However, for non-local treatment, solutions of the set of equations of correlations (\ref{eq38})--(\ref{eq41}) become non-trivial, which is a set of 4 (anisotropic convection) or 3 (isotropic approximation) second-order differential equations, and boundary conditions at both the surface and bottom of convective zone will be needed. Therefore, for non-local convection, it is to solve for boundary value problem of a set of 12 (anisotropic) or 10 (isotropic) equations. In our practice, Henyey method (Henyey et al. 1964) has been adopted to integrate the equations. Fig.~\ref{fig1} presents the auto-correlations of turbulent velocity and temperature fluctuations $x^2$ and $Z$, and their cross-correlation $V$ as functions of logarithmic gas pressure $\log P$, for non-local solar convective envelope in approximation of isotropic convection (the dashed lines) and the corresponding local model (the dotted lines) having the same depth of convective zone. As shown in Fig.~\ref{fig1}, both local and non-local models have virtually the same $x$, $Z$ and $V$ within the convectively unstable zone, revealing the fact that local convection theory is rather a very good first-order approximation. However, the two cases have distinct behaviors outside the convectively unstable region. Convective motions under local treatment are suddenly halted at the boundaries, while that in non-local theory $x$, $Z$ and $V$ reduce as a power law of $P$ in the overshooting zone just as expected as convective overshooting in the theory. Such a predicted trend has been proved by observations of the velocity and temperature fields of solar atmospheres, and also verified by hydrodynamic simulations (Deng et al. 2006).

\begin{figure}
\includegraphics[width=\columnwidth]{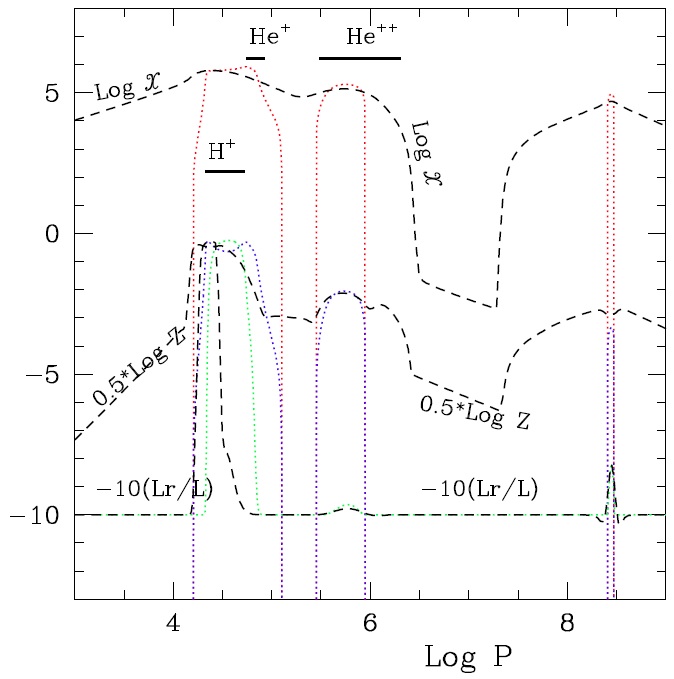}
\caption{$\log x$, $\log Z$, and $-10\left(L_r/L\right)$ versus $\log P$ for a model of $\delta$ Scuti star ($M=1.8M_\odot$,  $\log L/L_\odot=1.1948$ and $\log T_\mathrm{e}=3.8499$) calculated using local (dotted lines) and non-local (dashed lines) treatment.}\label{fig2}
\end{figure}

Convective motion properties shown in Fig.~\ref{fig1} actually represent general properties of the turbulent velocity and temperature fields of low-temperature stars with extended convective envelopes. For stars of somewhat higher temperatures, however, the structure of turbulent velocity and temperature fields becomes very different for the local and non-local convection models. Fig.~\ref{fig2} depicts the same quantities describing turbulent motions as in Fig.~\ref{fig1} but for a warm $\delta$ Scuti star ($M=1.8M_\odot$, $\log L/L_\odot=1.1948$, $\log T_\mathrm{e}=3.8499$), the dotted lines are for local models, and dashed lines for non-local. Also presented is the fractional radiation flux $L_r/L$ as a function of depth in terms of $\log P$. In Fig.~\ref{fig2}, horizontal bars denote respectively the partially ionized (5\%--95\%) zones of ionized hydrogen (H$^+$), the 1st and 2nd ionized helium (He$^+$, He$^{++}$). It is clear from Fig.~\ref{fig2} that, in the local model the 2nd ionized helium convective zone is completely detached from those of the ionized hydrogen and 1st ionized helium; whereas they are connected and form a single and larger convective zone in the non-local model due to overshooting from the zones. Such results enable plausible explanations to problems such as lithium depletions in the Sun and F-type stars, chemical abundance abnormals in the atmospheres of Ap stars, as well as stellar oscillation stabilities in general.

Diffusion induced by radiative acceleration and gravitational settling are usually adopted for the abundance anomaly for Ap stars (Richer \& Michaud 1993; Michaud \& Beaudet 1995). However, the diffusion time scale is too short in hot stars like Aps that have too thin convective zones. In this regard, some unknown turbulent diffusion enforcing an inverse mixing to extend the time scale (Richard et al. 2005). By applying our non-local convection theory in handling such a process, as demonstrated in Fig.~\ref{fig2}, the convective region in these hot stars is largely extended due to non-local convective diffusion. As a result, the diffusion time scale is naturally extended, therefore the manually added unknown turbulent diffusion is not any more needed.

\begin{figure}
\includegraphics[width=\columnwidth]{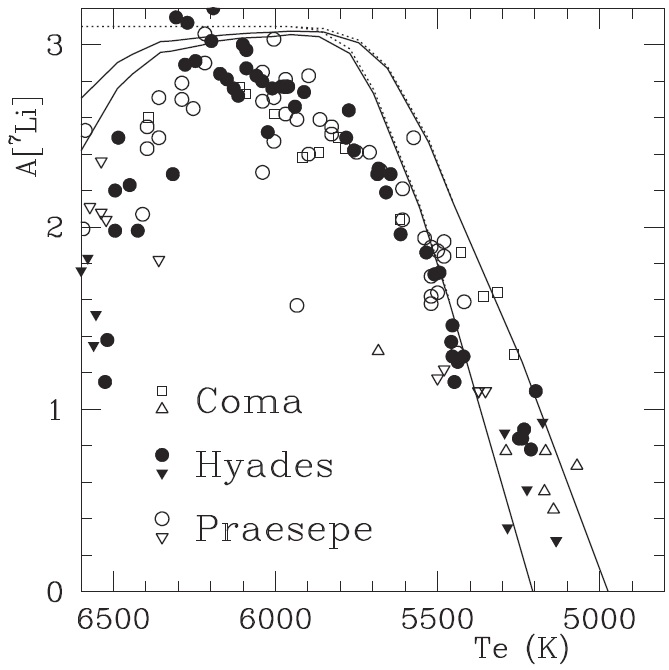}
\caption{Lithium abundance versus effective temperature for late-type dwarfs in Galactic open clusters Coma (of age 0.4--0.5Gyr), Hyades and Prasepe  (0.6--0.7Gyr). The solid and empty triangles represent the upper limits. The two solid lines are the theoretical isochrones at the ages of 0.4 and 0.7Gyr respectively.}\label{fig3}
\end{figure}

We have also calculated lithium depletion in late-type dwarfs (Xiong \& Deng 2009) using the same theory and parameter set, the results are presented in Fig.~\ref{fig3}. Two isochrones of age 0.4 and 0.7 Gyr were presented calibrated by using the Sun, the solid lines are models with gravitational settling, while the dotted lines without settling. The solid dot, circle and square symbols in Fig.~\ref{fig3}  represent respectively the observed atmospheric lithium abundance of member stars in three intermediate age open clusters, Coma (0.4--0.5Gyr), Hyades and Praesepe (0.6--0.7Gyr), the solid and empty triangles are the upper limits of lithium abundance from observations. Our theoretical predictions (the solid lines) agree pretty well with observational data in the three clusters. Towards the low temperature side, the depletion increases due to deepening of convective zone and more extended overshooting zone for lower temperatures; whereas at the high temperature end, higher depletions are caused by shorter gravitational settling time scale in stars with shallower convective zones. Presumably, our theory is also able to deal with abundance anomaly in Ap stars by implementing radiative acceleration in the scheme.

\begin{figure}
\includegraphics[width=\columnwidth]{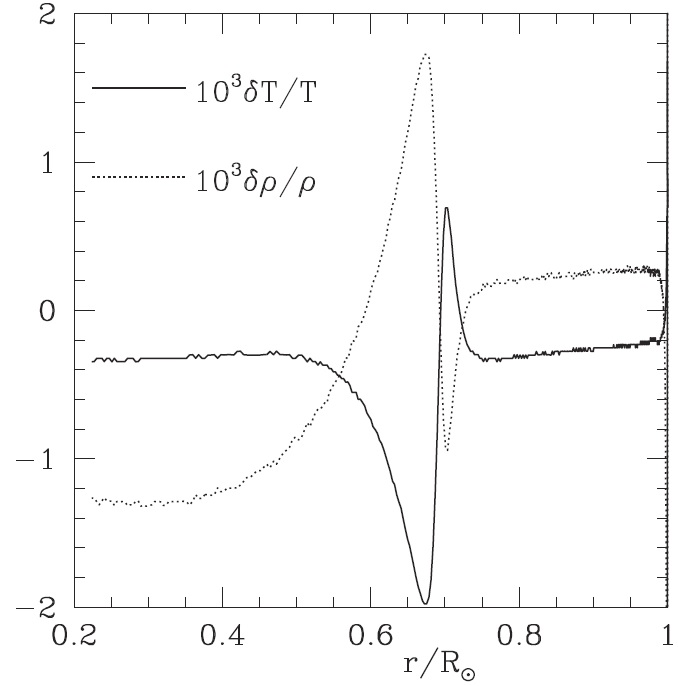}
\caption{The relative difference in temperature and density ($\delta T/T$ and $\delta\rho/\rho$) between the anisotropic and isotropic convection models as functions of fractional radius $r/R_\odot$.}\label{fig4}
\end{figure}

Theoretically, a complete non-local and anisotropic theory can better handle non-locality and isotropy of turbulent convection. A full set of equations of equilibrium structure in non-local and anisotropic theory of convection, however, may confront special complexity and are extremely difficult to integrate. Equations (\ref{eq30})--(\ref{eq33}) possess a certain type of singularity. Due to stability of numerical calculations, the radius of convergence for the integration is fairly small, and it needs special skill or maned operation to numerically calculate the integrations. This restricts application of the theory in general stellar physics research. When anisotropy of turbulent convection and overshooting mixing are not the major concern, eg. studying the temperature-pressure ($T-P$) structure of stars, isotropic convection is still a good approximation. Neglecting all terms containing $\chi^{ij}$ in equations (\ref{eq33})--(\ref{eq41}), equation (\ref{eq39}) cancels out, and the whole set of equations degrade two orders, and the number of equations in isotropic convection case is now 10. The working equations and boundary conditions in this situation can be found in our previous work (Xiong \& Deng 2001). In fact, the $T-P$ structures under isotropic and anisotropy conditions of non-local convection theory are quite similar. In Fig.~\ref{fig4}, we present relative difference in temperature $\delta T/T$ and density $\delta\rho/\rho$ between isotropic and anisotropy models with the same depth of convective zone of solar convective envelope as functions of depth ($\log P$). It is shown that they are rather close, with the maximum difference being 0.2 percent or smaller. Under the assumption of isotropic non-local convection, $\chi^{ij}=0$. Neglecting non-local convective diffusion term in equation (\ref{eq39}), and setting $Q^{ij}\nabla_j=0$, for equilibrium stellar model we have the following approximation,

\bq
\chi^{11}\approx \frac{\alpha c_1\bar P V}{\sqrt{3}\eta_e\bar\rho x\left(1+c_3\right)}.
\label{eq48}
\eq

\begin{figure}
\includegraphics[width=\columnwidth]{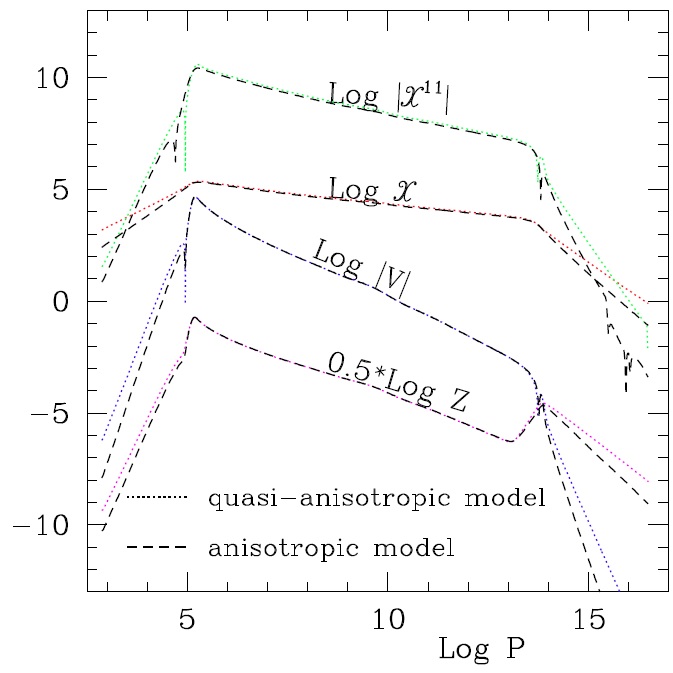}
\caption{$\log x$, $\log Z$, $\log |V|$ and $\log\chi^{11}$ versus $\log P$ for the anisotropic (dashed lines) and quasi-anisotropic (dotted lines) convection models of the Sun.}\label{fig5}
\end{figure}

Using isotropic and non-local convection theory, together with the quasi-anisotropic approximation in equation (\ref{eq48}), we have the numerical scheme of quasi-anisotropic and non-local convection ready for stellar models. Fig.~\ref{fig5} gives the correlations of turbulent velocity and temperature fluctuations $x^2$, $Z$, $V$ and $\chi^{11}$ as functions of depth ($\log P$) of quasi-isotropic (the dotted lines) and anisotropic (dashed lines) models of the Sun. It is clearly shown in Figs.~\ref{fig4} and \ref{fig5} that, in terms of not only the $T-P$ structure, but also turbulent velocity and temperature fields, the quasi-anisotropic and the full anisotropic models nearly resemble each other. In practice, isotropic modeling are far better in terms of convergence and stability in the computation than full anisotropic treatment. Taking such an advantage, we usually consider quasi-anisotropic model as a good replacement of the full anisotropic one for the calculations the equilibrium models. As we will show in Section~\ref{sec4}, the results of oscillation models based on the two approaches are nearly identical.

\section{Non-adiabatic oscillations of stars}\label{sec4}

\subsection{Radial and non-radial non-adiabatic oscillations of stars}\label{sec41}

The calculations of equilibrium models of stars in non-local and anisotropic convection theory have been described in Section~\ref{sec3}, which is the starting point of linear adiabatic and non-adiabatic oscillations modeling. When stars oscillate in small amplitudes near the equilibrium, oscillation quantities of the following are introduced,

\bq
y_i=\frac{\delta X_i}{X_i}e^{i\omega t}=Y_ie^{i\omega t},
\label{eq49}
\eq

\noindent where $Y_i$ is the eigen amplitude of oscillations, $\omega$ is its eigen frequency. For adiabatic stellar oscillations, $Y_i$ and $\omega$ are both real numbers; whereas for non-adiabatic oscillations they are complex numbers, the real and imaginary parts of $Y_i$ represent respectively the amplitudes and phase shifts of oscillations, and those of $\omega$ represent frequencies and amplitude growth rates of oscillations. Inserting equation (\ref{eq49}) into the dynamic equations for stellar structure and oscillations, equations (\ref{eq33})--(\ref{eq41}), and making Taylor expansion, retaining only the first-order terms, we will have the equations of linear non-adiabatic oscillations. For radial oscillations of stars, the Poisson equation describing self-gravitation field of stars equation (\ref{eq37}) has already initial integration, $\partial\phi/\partial r=GM_r/r^2$, therefore can be left out. Velocity $u^i$ and velocity-temperature correlation $V^i$ both have only radial components, therefore the number of equations for linear radial oscillations is 10. The working equations and boundary conditions can be found in our earlier work (Xiong et al. 1998b; Xiong \& Deng 2007). In literature, the so-called troublesome spatial oscillations of thermodynamic variables were reported when applying the local time-dependent theory of convection to deal with the coupling between convection in non-adiabatic oscillations of stars, and stellar oscillation in general (Keeley 1977; Baker \& Gough 1979; Gonczi \& Osaki 1980). By applying non-local time-dependent theory of convection, and careful assessments of the boundary conditions (Xiong et al. 1998a), such spatial oscillations of the thermodynamic variables are largely reduced or even completely disappeared in some cases.

For non-radial oscillations in stars, convection causes transportation of energy and momentum not only in radial direction, but also in transverse directions. The phenomenological time-dependent theory of convection can hardly handle the coupling between convection and oscillations in an accurate way. In equations (\ref{eq31})--(\ref{eq41}), $V^i$ is a three dimensional vector, and $\chi^{ij}$ is a second-order tensor, both are originated  from hydrodynamic equations. Our theory therefore has very solid foundation of hydrodynamics, and we anticipate that equations (\ref{eq38})--(\ref{eq41}) will express dynamic behaviors of turbulent convection in a more sounded way. When both rotation and magnetic field are neglected for single stars, motions in spherical coordinate system ($r$, $\theta$, $\phi$) should possess rotational symmetry with respect to radial direction. Therefore the vectors, $u^i$ and $V^i$, will have two independent components ($u_r$, $u_h$) and ($V_r$, $V_h$). For the equilibrium model and radial oscillations, all the non-diagonal components of $\chi^{ij}$ vanish,

\bq
\chi^{ij}=0,\; \; \; \textrm{when}\;\; i\not = j,
\label{eq50}
\eq

\noindent and it is clear from equation (\ref{eq29}) that $\chi^{ij}$ has only one independent component in this case,

\bq
\chi_2^2=\chi_3^3=-\chi_1^1/2.
\label{eq51}
\eq

However, for non-radial oscillations of stars, equation (\ref{eq50}) does not hold. It is extremely difficult to handle $\chi^{ij}$ in a precise way. Fortunately, apart from the central most parts, nearly all oscillations in the outer layers of stars oscillate dominantly in the radial direction (Unno et al. 1989), therefore equations (\ref{eq50}) and (\ref{eq51}) still hold approximately and $\chi^{ij}$ has only one independent component. Given such conditions, the order of the full set of equations for non-radial oscillations becomes 16.

Using the stellar evolutionary code from Padova group (Bressan et al. 1993), we calculated totally 14 evolutionary sequences for stellar mass $M=0.6-3.0M_\odot$ with the solar abundance ($X=0.70, Z=0.02$) and an intermediate overshooting ($l_\mathrm{over}=0.5H_P$). Taking luminosities and effective temperatures ($\log L/L_\odot$, $\log T_\mathrm{e}$) along the evolutionary tracks, we calculated 6 series of models of convective envelope, using working equations presented in previous section. Where series a is that of the completely non-local and anisotropic convection models, whereas series b--f are all for quasi-anisotropic convection models. In series b--f, non-local and isotropic convection models are calculated a prior, then the anisotropic component of turbulence, $\chi^{11}$, is derived using the quasi-anisotropic approximation equation (\ref{eq48}). The parameters used for all 6 series of models are given in Table~\ref{tb1}, in which the first column is the series number, the second column shows how the models are calculated: C.A. for completely anisotropic, Q.A. for quasi-anisotropic, and the third to fifth columns give the convection parameters used in the calculations of the corresponding series.

\begin{table}
\caption{Model parameters}\label{tb1}
\begin{tabular}{ccccc}
\hline
Series & type & $c_1$ &$c_2$&$c_3$\\
\hline
a & C.A. & 0.64&0.32&3\\
b & Q.A. & 0.70&0.35&3\\
c & Q.A. & 0.60&0.30&3\\
d & Q.A. & 0.78&0.39&3\\
e & Q.A. & 0.70&0.35&2\\
f & Q.A. & 0.70&0.35&5\\
\hline
\end{tabular}
\end{table}

\begin{figure}
\includegraphics[width=\columnwidth]{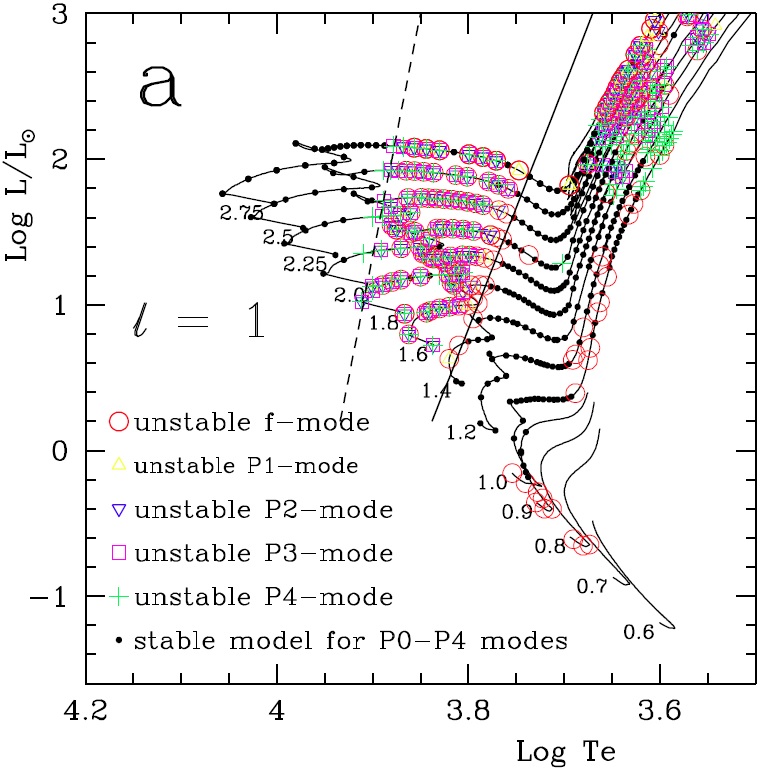}
\includegraphics[width=\columnwidth]{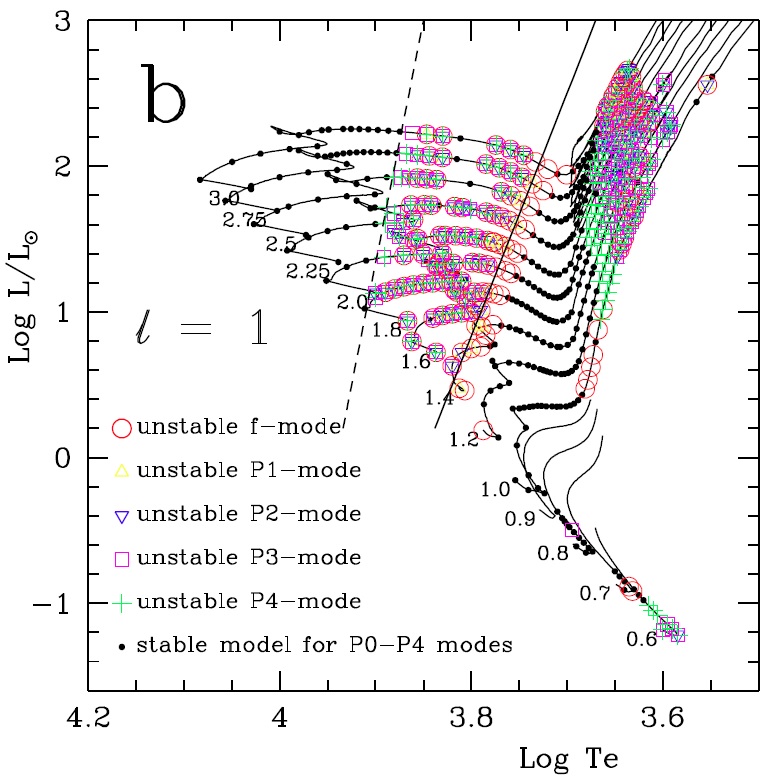}
\caption{H--R diagram for instability. Pulsationally stable in F--p4 modes (small dotted) and unstable in F (circles), p1 (open triangles), p2 (open inverse triangles), p3 (open squares) and p4 (plus signs) on the evolutionary tracks calculated for $M=0.6-3.0M_\odot$. Panel a): models in complete anisotropic equilibrium models; Panel b): models in quasi-anisotropic equilibrium models. The dashed and solid lines are respectively the theoretical blue and red edges of instability strip for complete anisotropic models.}\label{fig6}
\end{figure}

Given the parameters of the series, we calculated non-adiabatic oscillations in low-degree ($l=1-4$) g9--p29 modes for series a--f defined in Table~\ref{tb1}. All the pulsationally stable and unstable modes (F--p4) of series a and b are presented in Fig.~\ref{fig6}a and \ref{fig6}b, where the circles, triangles, inverse triangles and plus signs are used to denote unstable (F--p4) modes, all stable modes are in small dots. From Fig.~\ref{fig6}, one can make it clear that, the stabilities in the two series are similar but the red edge of $\delta$ Scuti strip for the series b seems redder compared with series a. This infers that quasi-anisotropic calculations approximate the complete anisotropic convection theory perfectly in terms of both equilibrium model and stability of linear oscillation calculations. Therefore, the quasi-anisotropic approximation is a very useful and accurate approach.

\subsection{Dependence of pulsation stability on the convective parameters}\label{sec42}

Many researches on stellar convection demonstrate in one way or the other that the ``mixing-length'' used in different kinds of theories is not a fixed parameter, which varies slowly as a function of stellar parameters such as mass, luminosity and effective temperature etc. (Ludwig et al. 1999). The conventional way of calibration is to use solar observations, but the old question is still there: can it be used to stars other than the Sun? If the pulsation stability depends sensitively on the choice of the convective parameters, the results of the theoretical calculations of the pulsation stability are questionable.

\begin{figure*}
\includegraphics[width=15cm]{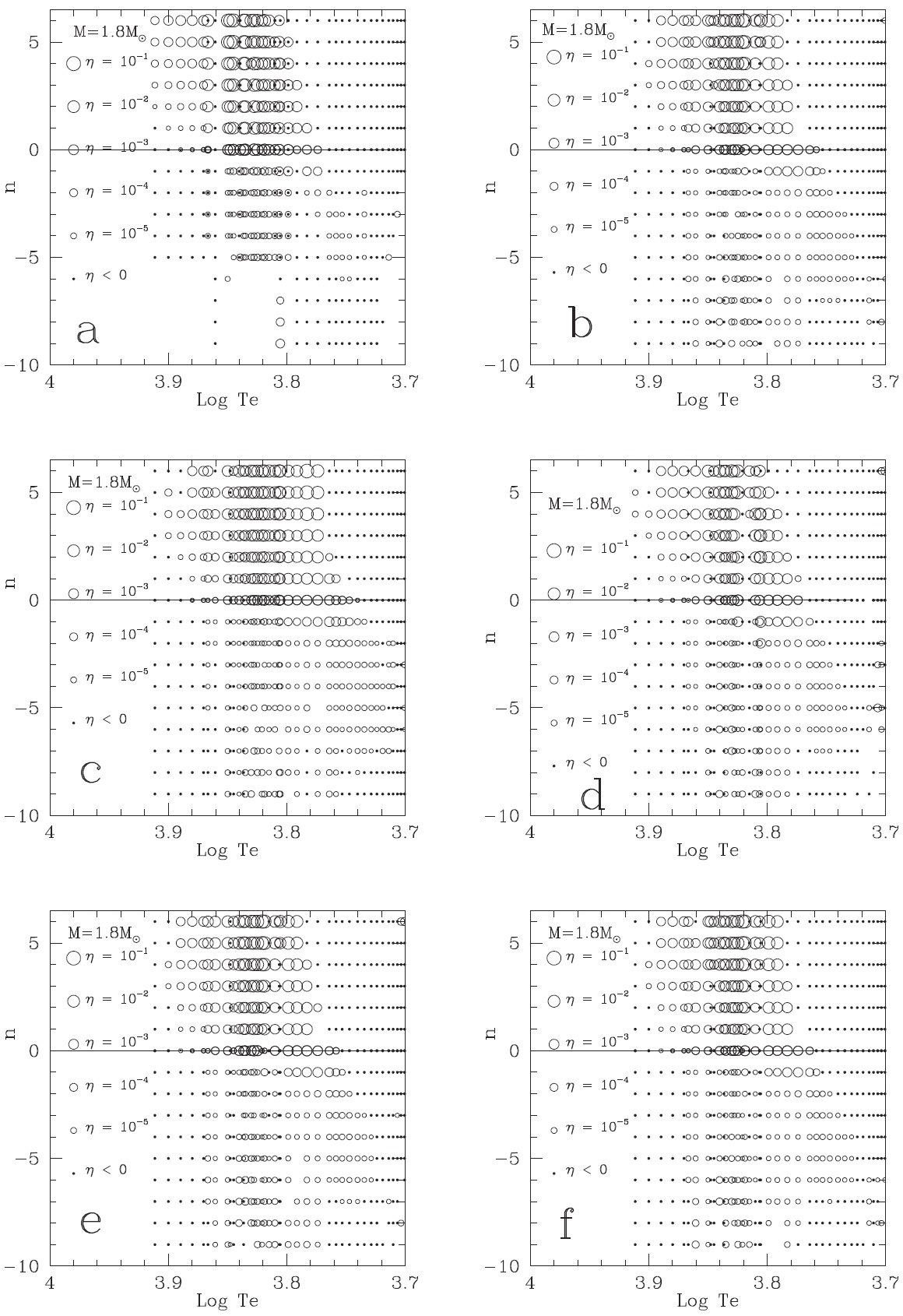}
\caption{Pulsationally stable (small dots) and unstable (circles) modes in $n$--$\log T_\mathrm{e}$ plane for models along the evolutionary track of a $M=1.8M_\odot$ star. $n$ is the radial order of oscillation and $T_\mathrm{e}$ is the effective temperature of the stellar model. Panel a): the anisotropic equilibrium model; Panels b)--f): the quasi-anisotropic models with parameters given in Table~\ref{tb1}, see text for details.}\label{fig7}
\end{figure*}

In order to investigate the dependence of pulsation instability on convection parameters, we use different combinations of convection parameters in Table~\ref{tb1} for linear non-adiabatic calculations (series b--f). The results of theoretical stability analysis in g9--p5 modes for a $M=18.M_\odot$ star using all the parameter combinations in Table~\ref{tb1} (series a--f) in evolutionary stages between MS until SGB ($\log T_\mathrm{e} > 3.70$) are presented in the 6 panels of Fig.~\ref{fig7}. In all the panels of Fig.~\ref{fig7}, the horizontal axis is stellar effective temperature ($\log T_\mathrm{e}$), and the vertical axis is the radial orders of oscillation modes ($n_r$ runs in between -9 to 5, correspond to g9--p5 modes), the horizontal thin solid line on each panel is F-mode ($n_r=0$), above which are all p-modes, and below g-modes. The small dots denote all stable models, whose amplitude growth rates are negative ($\eta=-2\pi\omega_i/\omega_r<0$), where $\omega_i$ and $\omega_r$ are respectively the imaginary and real parts of the complex frequency of oscillations ($\omega=i\omega_i+\omega_r$); whereas the circles are the pulsationally unstable models, whose size is proportional to $\log\eta$.  It is clear from Fig.~\ref{fig7}a and \ref{fig7}b that if one compares the models in series a and b, all the p-modes are nearly identical, and the g-modes are different but similar. This again demonstrates that the quasi-anisotropic approximation is a good approach.

Models in b, c and d series are all calculated using quasi-anisotropic convection having the same $c_2/c_1=0.5$ and $c_3=3$, but different $c_1$. As presented in Fig.~\ref{fig7}b--\ref{fig7}d that the three sets of results are extremely close to each other. Although $c_1$ varies by 30\% in these models, the red-edges of the pulsational instability trip is shifted to high temperatures by only 5\% (~$\Delta\log T_\mathrm{red}\sim 0.02$) when $c_1$ goes from 0.70 to 0.95.

Fig.~\ref{fig7}b, \ref{fig7}e and \ref{fig7}f represent 3 types of quasi-anisotropic convection models that possess the same $c_1$ and $ c_2$ ($c_1=2c_2=0.85$) but different $c_3$ varying from 2 to 5. The degree of anisotropy in such a range of $c_3$ changes greatly, with the r.m.s turbulent velocity ratio, $\overline{w'^2_r}/\overline{w'^2_h}\approx\left(3+c_3\right)/2c_3$, changes from 1.25 to 0.8. However the pulsation stability does not respond sizably in such a large variation range of $c_3$, this implies that the pulsation stability does not depend, at least not sensitively, on the choice of parameter of anisotropy $c_3$.

\newpage
\section{Summary and conclusions}\label{sec5}

In this paper, we present a dynamic theory for complete non-local and anisotropic stellar convection, and the computational results on the structure of stellar envelope structure and non-adiabatic oscillations based on such a theory. This work can be summarized in the following,

\begin{enumerate}
\item Starting from the first principle of hydrodynamic equations and applying Reynold's decomposition for turbulence, we established a set of dynamic equations for mean fluid, and that of auto- and cross-correlations of turbulent velocity and temperature fluctuations. Following the theory of turbulence, we have introduced a proper way to deal with dissipation, diffusion and anisotropy of turbulent  convection.  A self-consistent and closed formalism of turbulent convection is developed, which is ready for the calculations of stellar structure and oscillations.

\item Three convection parameters ($c_1$, $c_2$, and $c_3$) are included in the theory, which are respectively connected to dissipation, diffusion and anisotropy of turbulent convection in stars.

\item Discussions were made on the calculations of the equilibrium envelope model and the calibration of the convective parameters. Numerical experiments suggested that ($c_1$, $c_2$, $c_3$)=(0.64, 1/2, 3) is a robust choice, which has been justified by a number of relevant observational facts including: the predicted structure of solar convective envelope that agrees with solar seismic inversion, the structures of atmospheric turbulent velocity and temperature fields of the Sun that match solar observations, and the depletions of lithium in the Sun and solar-type stars.

\item When the anisotropy of turbulent velocity and overshooting mixing in stars are not the major concerns, isotropic treatment of convection is a fairly good approximation that gives good enough $T-P$ structure of stars. Only two parameters $c_1$ and $c_2$ are required in the theory. Our results show that ($c_1$, $c_2$)=(0.70, 0.35) is a good set. The quasi-anisotropy approximation eq.~(\ref{eq47}) can deliver very good measurements of the anisotropic component $\chi^{11}$ for turbulent convection, therefore can be used for linear non-adiabatic oscillation analysis for the sake of computational effort.

\item The theory can be used to study both radial and non-radial oscillations of stars, and it handle both thermodynamic and dynamic couplings between convection and oscillations.

\item Using the scheme of non-adiabatic oscillations in the non-local and time-dependent convection theory, we have done linear stability analysis of low degree ($l=1-4$) g9--p29 modes for stars with $M=0.6-3.0M_\odot$ from MS to RGB, and up to AGB phase. Theoretical calculations clearly show two separate $\delta$ Scuti and Mira-like instability strips in the H-R diagram. For future work, we will carry out further theoretical analysis of $\delta$ Scuti and $\gamma$ Doradus stars and pulsating red giant stars, and compare the results with observations.

\item The dependence of pulsation stability of stars on the convection parameters ($c_1$, $c_2$, $c_3$) was studied very carefully. The results show that, the stability of oscillations has only weak response to the choice of the 3 convection parameters within a rather wide range.

\end{enumerate}

\section*{Acknowledgements}
This work was supported by National Natural Science Foundation of China (NSFC) through grants 11373069, 11473037, and 11403039.




\begin{thebibliography}{99}

\bibitem[\protect\citeauthoryear{Baker \& Gough}{1979}]{BG79}Baker, N. H., Gough, D. O. 1979, \apj, 234, 232

\bibitem[\protect\citeauthoryear{B\"ohm-Vitense}{1958}]{BV58}B\"ohm-Vitense, E. 1958, Z. Ap., 46, 108

\bibitem[\protect\citeauthoryear{Bressan et al.}{1993}]{Brs93}Bressan, A., Faggotto, F., Bertelli, G., Chiosi, C. 1993, \aaps, 100, 647

\bibitem[\protect\citeauthoryear{Canuto}{1993}]{Ca1993}Canuto, V. M. 1993, ApJ, 416, 331

\bibitem[\protect\citeauthoryear{Cox}{1980}]{Cox80}Cox, J. P. 1980, Theory of Stellar Pulsation (Princeton: Princeton Univ. Press)

\bibitem[\protect\citeauthoryear{Deng et al.}{2006}]{DXC06}Deng, L., Xiong, D. R., Chan, K. L. 2006, \apj, 643, 426

\bibitem[\protect\citeauthoryear{Deng \& Xiong}{2008}]{DX08}Deng, L. Xiong, D.R. 2008, \mnras, 386, 1979

\bibitem[\protect\citeauthoryear{Gonczi \& Osaki}{1980}]{GO80}Gonczi, G., Osaki, Y. 1980, \aap, 84, 304

\bibitem[\protect\citeauthoryear{Gough}{1977}]{Go77}]Gough, D. O. 1977, \apj, 214, 196

\bibitem[\protect\citeauthoryear{Grigahc\`ene et al.}{2005}]{GDG05}Grigahc\`ene, A., Dupret, M.-A., Gabriel, M., Garrido, R., Scuflaire, R. 2005, \aap, 434, 1055

\bibitem[\protect\citeauthoryear{Grossman}{1996}]{Gr96}Grossman, S. A. 1996, \mnras, 279, 305

\bibitem[\protect\citeauthoryear{Henyey et al.}{1964}]{HFG64}Henyey, L. G., Forbes, J. E., Gould, N. L. 1964, \apj, 139, 306

\bibitem[\protect\citeauthoryear{Hinze}{1975}]{H75}Hinze, J. O. 1975, Turbulence (New York: McGraw-Hill)

\bibitem[\protect\citeauthoryear{Keeley}{1977}]{Kl77}Keeley, D. A. 1977, \apj, 211, 926

\bibitem[\protect\citeauthoryear{Keil \& Canfield}{1978}]{KC78}Keil, S. L., Canfield, R. C. 1978, \aap, 70, 169

\bibitem[\protect\citeauthoryear{Komm et al.}{1991}]{KMN91}Komm, R., Mattig, W., \& Nesis, A. 1991, \aap, 243, 251

\bibitem[\protect\citeauthoryear{Ledoux \& Walraven}{1958}]{Ldx58}Ledoux, P., Walraven, T. 1958, in Handbuch der Physik Vol. 51, ed. S. Fl\"ugge (Berlin: Springer-Verlag), 353

\bibitem[\protect\citeauthoryear{Ludwig et al.}{1999}]{LFS99}Ludwig, H.-G., Freytag, B., Steffen, M. 1999, \aap, 346, 111

\bibitem[\protect\citeauthoryear{Michaud \& Beaudet}{1995}]{MB95}Michaud, G., Beaudet, G. 1995, Mem. Soc. Astron. It., 66, 477

\bibitem[\protect\citeauthoryear{Nesis \& Mattig}{1989}]{NM89}Nesis, A., Mattig, W. 1989, \aap, 221, 130

\bibitem[\protect\citeauthoryear{Richard et al.}{2005}]{RMR05}Richard, O., Michaud, G., Richer, J. 2005, \apj, 619, 538

\bibitem[\protect\citeauthoryear{Richer \& Michaud}{1993}]{RM93}Richer, J., Michaud, G. 1993, \apj, 416, 312

\bibitem[\protect\citeauthoryear{Rotta}{1951}]{Ro51}Rotta, J. C. 1951, Z. Phys., 129, 547

\bibitem[\protect\citeauthoryear{Spiegel}{1963}]{Sp63}Spiegel, E. A. 1963, \apj, 138, 216

\bibitem[\protect\citeauthoryear{Stellingwerf}{1982}]{St82}Stellingwerf, R. F. 1982, \apj, 262, 330

\bibitem[\protect\citeauthoryear{Ulrich}{1970}]{Ulr70}Ulrich, R. K. 1970, \apj, 162, 993

\bibitem[\protect\citeauthoryear{Unno}{1961}]{U63}Unno, W. 1961, PASJ, 13, 276

\bibitem[\protect\citeauthoryear{Unno}{1967}]{U67}Unno, W. 1967, PASJ, 19, 140

\bibitem[\protect\citeauthoryear{Unno et al.}{1989}]{UOASS89}Unno, W., Osaki, Y., Ando, H., Saio, H., Shibahashi, H. 1989, Nonradial Oscillations of Stars (Tokyo: Univ. of Tokyo Press).

\bibitem[\protect\citeauthoryear{Xiong}{1978}]{X78}Xiong, D. R. 1978, Chin. Astron., 2, 118

\bibitem[\protect\citeauthoryear{Xiong}{1989}]{X89}Xiong, D. R. 1989, \aap, 209, 126

\bibitem[\protect\citeauthoryear{Xiong \& Deng}{2001}]{XD01}Xiong, D. R., Deng, L. 2001, \mnras, 327, 1137

\bibitem[\protect\citeauthoryear{Xiong \& Deng}{2007}]{XD07}Xiong, D. R., Deng, L. 2007, \mnras, 378, 1270

\bibitem[\protect\citeauthoryear{Xiong \& Deng}{2009}]{XD09}Xiong, D. R., Deng, L. 2009, \mnras, 395, 2013

\bibitem[\protect\citeauthoryear{Xiong et al.}{1997}]{XCD97}Xiong, D. R., Cheng, Q. L., Deng, L. 1997, \apjs, 108, 529

\bibitem[\protect\citeauthoryear{Xiong et al.}{1998a}]{XCD98}Xiong, D. R., Cheng, Q. L., Deng, L. 1998a, \apj, 500, 449

\bibitem[\protect\citeauthoryear{Xiong et al.}{1998b}]{XDC98}Xiong, D. R., Deng, L., Cheng, Q. L. 1998b, \apj, 499, 355

\bibitem[\protect\citeauthoryear{Zhang et al.}{2012}]{ZDXC12}Zhang, C., Deng, L., Xiong, D.R., Christensen-Dalsgaard, J. 2012, \apjl, 759, L14

\end{thebibliography}



\bsp	
\label{lastpage}
\end{document}